\documentclass[aps, pra, a4paper,showpacs,twocolumn,10pt]{revtex4-1}
\usepackage{bbm, amsmath, amssymb, amsthm, bm,textcomp, nicefrac,geometry,ragged2e}
\usepackage{graphicx,epstopdf,color,verbatim,enumitem}
\usepackage[dvipsnames]{xcolor}
\usepackage[bbgreekl]{mathbbol}

\usepackage{float}
\usepackage{subfigure} 

\usepackage{multirow} 
\usepackage{tabularx} 

\newcolumntype{C}[1]{>{\centering\arraybackslash}p{#1}}

\begin{document}

\title{Entangle me!\\ A game to demonstrate the principles of Quantum Mechanics}

\author{Andrea L\'opez-Incera}
\email{andrea.lopez-incera@uibk.ac.at} 
\author{Wolfgang D\"ur}
\email{wolfgang.duer@uibk.ac.at}
\affiliation{Institut f\"ur Theoretische Physik und Institut f\"ur Fachdidaktik, Bereich DINGIM,
Universit\"at Innsbruck, Technikerstrasse 21a, A-6020 Innsbruck, Austria}

\begin{abstract}
We introduce a game to illustrate the principles of quantum mechanics using a qubit (or spin-first) approach, where students can experience and discover its puzzling features first-hand. Students take the role of particles and scientists. Scientists unravel underlying rules and properties by collecting and analysing data that is generated by observing particles that act according to given rules. We show how this allows one to illustrate quantum states, their stochastic behavior under measurements as well as quantum entanglement. In addition, we use this approach to illustrate and discuss decoherence, and a modern application of quantum features, namely quantum cryptography. We have tested the game in class and report on the results we obtained.
\end{abstract}

\maketitle 

\section{Introduction}
Quantum mechanics is a subject that is notoriously difficult to teach in class. In addition to the conceptual difficulties that are intrinsically connected to its counter-intuitive features, an in-depth understanding requires advanced mathematics. In contrast to other areas of physics, it is hard to find contexts from everyday experience, making it difficult to provide motivating or illustrating examples. Furthermore, hands on experiments, or even demonstration experiments, are unavailable or seriously limited. Perhaps for these reasons, concepts to teach quantum mechanics at high-school level are far less numerous and less developed as in other areas of physics (see however \cite{For66, Hoo93, Mul02, Sch10, Nie03, Mer03, Hob07, Pos08, Car09, Heu10, Win13, Dur13, Dur14, Koh14, Koh15, Dur16, Qlab}).

Here we suggest a game to illustrate the principles of quantum mechanics (see also \cite{Cor18} for an alternative approach), where students can experience first-hand how the rules of quantum mechanics work, and what they imply for the properties of systems. Students will play the role of both scientists and quantum particles, simulating a real laboratory. This allows them not only to play and behave like scientists, but helps them to internalize the non-classical features and strange properties of a quantum world. The role-play in science teaching \cite{McSha00} has attracted interest over the past years, and there exist several examples \cite{Beg04, Sin10, Bru16} in the literature of different applications of kinesthetic activities used to teach concepts of physics. Kinesthetic activities provide direct illustrations of the physical concepts, which makes it easier for the students to create image schemas that help understanding. At the same time, these type of activities enhance motivation and enjoyment. In addition, the game we propose gives the student the opportunity of obtaining the same experimental results that they would obtain in a real laboratory.

The game we introduce has three parts, where (i) quantum superposition states (together with Heisenberg's uncertainty relation); (ii) quantum entanglement and (iii) decoherence are illustrated. This covers more than is often taught in introductory courses on quantum mechanics at undergraduate level. Nevertheless, we believe that with our approach these three fundamental principles which are at the core of quantum mechanics can be illustrated and experienced. With the same rules, also quantum cryptography \cite{Gis07, Gis02, Dur13} can be explored in a game-like fashion.

We stress that no previous mathematical background is needed. In fact, it is even desirable that the students do not have any previous knowledge of quantum mechanics, so that they can come up with fresh and imaginative theories after analyzing their own measurement results.

\subsection{Theoretical background}\label{theorback}

We make use of the qubit (or spin first) approach to quantum mechanics, \cite{Man11, Dur13, Dur14, Dur16} where the simplest quantum mechanical system, a two-level system or qubit, is used to demonstrate the basic features of quantum theory. In contrast to a classical two-level system, a qubit can be in a superposition of two states, 
\begin{equation}\label{sup state}
|\psi\rangle = \alpha |0\rangle + \beta |1\rangle,
\end{equation}
where $\alpha$ and $\beta$ are complex numbers. Thus, there exist infinitely many superposition states. In this work, we concentrate only on four of these states to reduce the complexity of the game and make it understandable and tractable for the students. In particular, in addition to the (classical) basis states $\{|0\rangle ,|1\rangle \}$, we work with superposition states of the form,\cite{Dur13,Dur14}
\begin{equation}
|\pm \rangle =\frac{1}{\sqrt{2}}(|0\rangle \pm |1\rangle).
\end{equation}
If one associates states e.g. with two different positions, this means that such a superposition state describes a situation where the system is essentially at both places simultaneously. For the polarization degree of freedom of a single photon, the states $|0\rangle,|1\rangle$ correspond to horizontal and vertical polarization, while the superposition states $|\pm\rangle$ are $\pm 45{}^\circ $ polarized. The Bloch sphere is nicely suited to illustrate a qubit: \cite{Dur13,Dur14} standard basis states are represented by unit vectors pointing in $\pm z$ direction, while superposition states point in $\pm x$ direction. Notice that orthogonal vectors are anti-parallel in this picture. 

Measurements may be illustrated by a slit oriented in a certain direction (e.g. $z$ for $z$-property). \cite{Dur13,Dur14} Within this picture, measurement eigenstates \cite{eigenstates} are easy to visualize, since they are unit vectors pointing in the measurement direction, i.e. they can pass unaltered through the slit, giving the corresponding outcome deterministically. For instance, if one performs a measurement in the $z$ direction, the state $|0\rangle$ passes unaltered through the slit, giving the outcome $+1$, while a superposition state $|+\rangle$ pointing in $x$-direction has to flip up or down to pass. Thus, the outcome one gets when the state $|+\rangle$ is measured is random and cannot be predicted. In addition, the state of the qubit is changed –-the vector points in $+z$ or $-z$ direction afterwards. Measurements of different properties are possible, e.g. the $x-$property, where the slit is oriented in $x$-direction. Crucially, a quantum system cannot possess a deterministic $z$ and $x$ property simultaneously –-the properties are complementary--, which is the basis of Heisenberg's uncertainty relation.\cite{Dur13}

Apart from the states $|0\rangle,|1\rangle,|+\rangle,|-\rangle$ (and all the possible superpositions given by eq.~\eqref{sup state}) described above --called pure states--, there can also exist probabilistic mixtures of pure states, called mixed states. As an example, let us consider an ensemble of $N$ qubits, that are either in the $|0\rangle$ or the $|1\rangle$ state with equal probability. This mixed state is represented in the Bloch as the null vector. If one performs a series of measurements on the $N$ qubits, the outcomes will be random in whatever direction one measures. However, a pure state of $N$ qubits, e.g. all in the state $|0\rangle$, will always give the outcome $+1$ when measured in the $z$-direction.

For systems of two qubits, the superposition principle leads to the possibility of entangled states, e.g. of the form $|\phi^+\rangle=(|0\rangle|0\rangle + |1\rangle |1\rangle)/\sqrt{2}$.\cite{tensor} Such states have the unique feature that measurements of specific properties on individual systems lead to random outcomes, but nevertheless the outcomes of the two systems are perfectly correlated. This is not only true for the $z$-property, but also for the $x$-property, as can be seen by noting that $|\phi^+\rangle=(|+\rangle|+\rangle + |-\rangle |-\rangle)/\sqrt{2}$. The latter is what makes this a unique quantum feature that does not exist in a classically system.

Finally, entanglement (with environmental particles) can also be seen as source of decoherence. \cite{Heu18} Decoherence \cite{Zurek,Zurek2} is a mechanism that allows one to explain the absence of quantum effects in large or poorly isolated systems, and is thus an essential building block to understand the power and limitations of quantum mechanics. Entanglement that is built up due to interaction of a system with its environment leads to random behavior of the system due to lack of control on the environment. \cite{Heu18} This is a different kind of randomness as for pure states that are in a superposition state, or for entangled systems. \cite{Dur13, Heu18}

\subsection{Learning objectives}

The game is divided in three parts, each one designed to teach a specific set of concepts about Quantum Mechanics in a hierarchical way, i.e. the complexity of the concepts increases from part 1 to part 2, and from part 2 to part 3. This structure gives the teacher the flexibility to choose either to first play all parts and then work with all the concepts, or to play and work with each part separately. 

The learning objectives of each part are the following:

\begin{enumerate}
   \item Single particles (Sec.~\ref{part 1}).
   \begin{itemize}
     \item Quantum superposition states.
     \item Preparation and measurement processes (including measurement in different bases) of single quantum particles.
   \end{itemize}
   \item Entanglement (Sec.~\ref{part 2}).
   \begin{itemize}
     \item Preparation and measurement processes of entangled pairs of particles.
     \item Difference between classical and quantum correlations.
   \end{itemize}
   \item Decoherence (Sec.~\ref{part 3}).
   \begin{itemize}
     \item Effect of decoherence on single-particle states (difference between pure and mixed states).
     \item Effect of decoherence on entangled states.
   \end{itemize}
   
\end{enumerate}

Furthermore, the game is designed so that the students can also learn to work and think as scientists, which is of fundamental importance not only for their future careers but also for their personal development as citizens. The game can be used as an active learning approach (see Sec.~\ref{implementation}) to teach Quantum Mechanics and enhance critical thinking. Among other skills, we highlight that students can learn to:
\begin{itemize}
  \item Perform measurements.
  \item Analyze measurement results from a critical perspective, identifying the differences between what they expected from their previous knowledge and what they obtain.
  \item Come up with explanations for the obtained results. 
  \item Predict results based on these explanations and test them afterwards.
\end{itemize}

In addition, we explain how the game can be extended to simulate Quantum Cryptography in Sec.~\ref{QCryp}. In Sec.~\ref{class}, we report the experience and conclusions of testing the game in class with 16-year high-school students.

\section{Implementation of the game}\label{implementation}

In this section, we explain how the game is played, giving more details for the specific parts in sections~\ref{part 1},~\ref{part 2} and~\ref{part 3}. The students are initially arranged in two groups: one playing the role of scientists and the other one playing the role of quantum particles. In part 3, a third group (the environment) is introduced to deal with decoherence.

The goal of the particles is to avoid being measured by the scientists. The measurement process is illustrated by the scientists trying to hit the particles with a ball, analogously to e.g. sending photons. The particles can only move along two lines painted on the floor, following the rules specified in each part (see Sec.~\ref{part 1},~\ref{part 2} and~\ref{part 3}).

The role of the scientists is to figure out rules (i.e. physical laws) that properly describe the behavior of particles by observation, i.e. by preparing particles in specific states and by performing measurements. They have a source available, where they can push one out of four buttons in part 1. Depending on the chosen button (there is a one-to-one correspondence to the four states $\{|0\rangle,|1\rangle,|+\rangle,|-\rangle\}$ listed above which however only the particles know), the particle starts in a certain state and begins to run along the lines to avoid the balls thrown by the scientists. Once the scientist hits a particle with the ball, he/she has to choose which property of the particle he/she wants to measure: either the leg (associated to $z$-measurements) \textit{or} the arm property (associated to $x$-measurements). Then, the particle tells the measurement outcome to the scientist. The scientists should perform as many measurements as possible and should take notes of their experiment results, in order to come up later with a description of what is happening. That is, the scientists should figure out the rules and make predictions for new experiments, which they then test. In part 2, the scientists discover a hidden button on the machine. When this button is pressed, two particles emerge together in an entangled state. Now, the scientists work in pairs, where each scientist measures one particle.

In the following sections, the detailed rules for the particles in each part are explained. These rules are explained to the particles separately (the scientists cannot know them) at the beginning of the game. In this way, the scientists can experience how it is to do research while they learn (active learning) and can develop a critical thinking. \cite{rules}

\subsection{Part 1: Single Particles}\label{part 1}
\begin{figure}[ht!]
\centering
\includegraphics[width=3.4in]{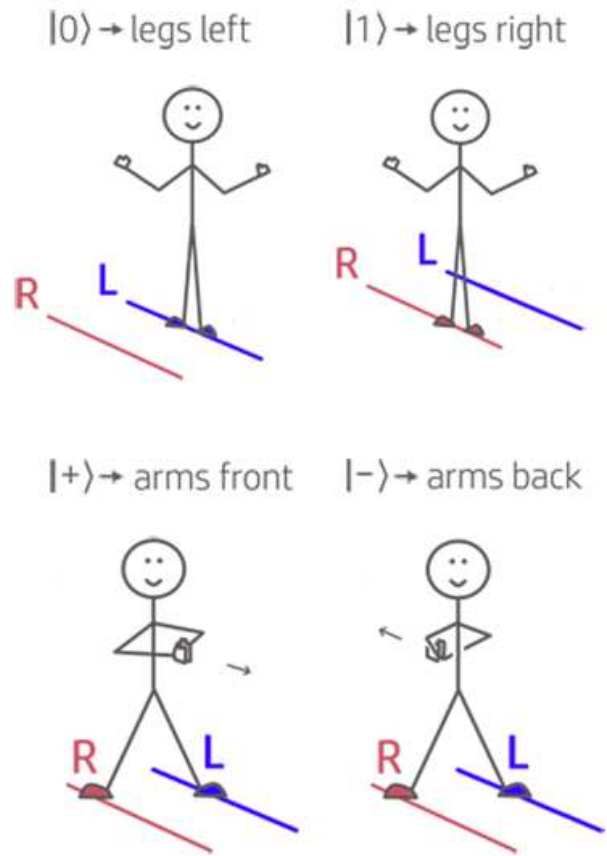}
\caption{Representation of the different states. For the state $|0\rangle$ ($|1\rangle$), the student stands on the left (right) line with both feet on it and outstretched arms. For the state $|+\rangle$ ($|-\rangle$), the student stands with one foot on each line and arms pointing forward (backward).}
\label{states}
\end{figure}

In order to play the role of quantum particles (qubits in our case), students should be able to stay not only in one of the two levels of the qubit, but also in a superposition state of both. To do so, we can paint two lines on the floor: one on the left and one on the right, and standing with both legs on one line represents the basis states $|0\rangle$ and $|1\rangle$. However, the students can be at the two lines at the same time by just putting a foot on each line, which represents the superposition state. In addition, particles act according to the following simple rules (that only they know), and that correspond to the actual behavior of quantum particles. They can be prepared in four different states, $\{|0\rangle,|1\rangle,|+\rangle,|-\rangle\}$ where they stand with both legs and outstretched arms on the left line (state $|0\rangle$) or right line (state $|1\rangle$), or with legs on different lines where arms point forward (state $|+\rangle$) or backward (state $|-\rangle$) (see Fig.~\ref{states}). \cite{alternative rep} Thus, the general rule for the particles in this part is that if they stand with separated legs, the arms should be together, and vice versa. Note that separated legs and outstretched arms represent that the corresponding property is unspecified.

The difference between pure and mixed states can also be addressed. Pure states have one property completely specified, i.e. either legs or arms together, leading to a deterministic outcome (always the same) if the corresponding property is measured. However, mixed states lead to random outcomes for whatever property one measures (see Sec.~\ref{theorback}). Since separated limbs represent an unspecified property and lead to random measurement outcomes, a mixed state can be represented by the student standing with outstretched arms and separated legs.

\subsubsection{Measurement process}

\begin{figure}[H]
\centering
\subfigure[$\,\,\,|+\rangle=\frac{1}{\sqrt{2}}(|0\rangle + |1\rangle)\rightarrow\,|0\rangle$\label{measa}]{\includegraphics[width=3.4in]{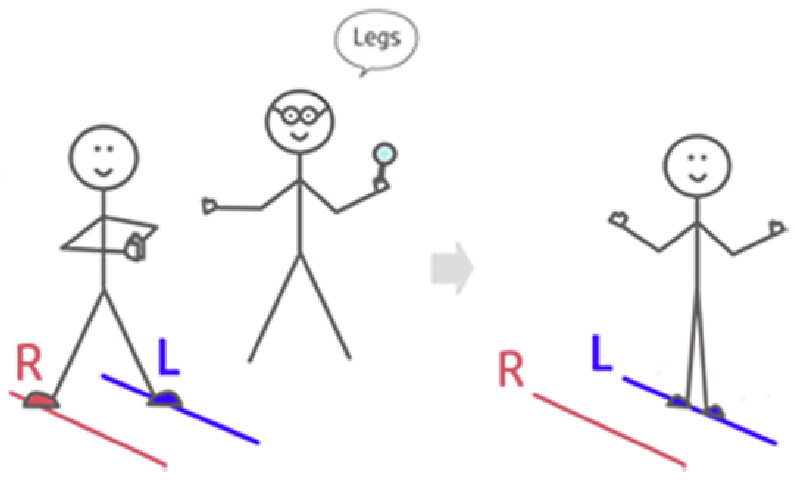}}
\subfigure[$\,\,\,|0\rangle=\frac{1}{\sqrt{2}}(|+\rangle + |-\rangle)\rightarrow\,|+\rangle$\label{measb}]{\includegraphics[width=3.4in]{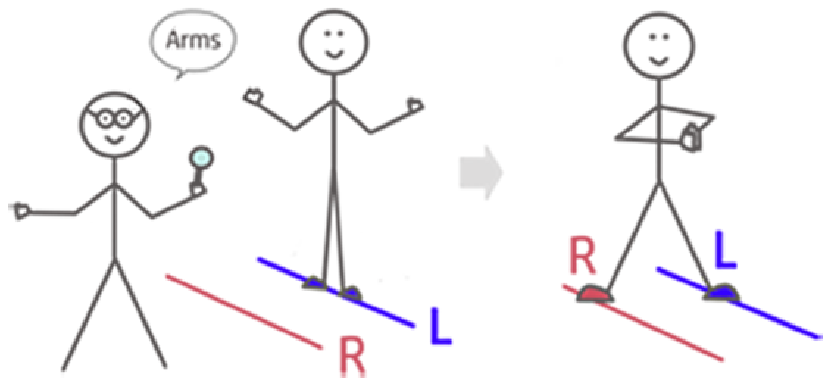}}
\caption{Measurement of the (a) \textit{leg} or (b) \textit{arm} property on superposition states. When the scientist measures a particle that is in a superposition state on the chosen property (basis), the particle has to randomly jump to either one state or the other of the superposition.}
\end{figure}

Once a particle is measured by a scientist (hit with the ball), the scientist has to choose between two different kinds of measurements: \textit{legs} (corresponding to $z$-measurement, with outcomes left and right), or \textit{arms} (corresponding to $x$-measurement, with outcomes front and back). If the \textit{leg} property is measured, and the particle stands on one line, the student announces the result (left line or right line). If the particle is standing on both lines, the student can choose to jump with the two feet to the left or right line randomly (outstretched arms now), and announces the result (Fig.~\ref{measa}). Similarly, if the \textit{arm} property is measured, the student tells if her/his arms are pointing forward (front) or backward (back). If the arms are outstretched, the student chooses randomly if her/his arms will point forward or backward, changing position so that the legs are on different lines (see Fig.~\ref{measb}), and announces the resulting state. 

\subsection{Part 2: Entanglement}\label{part 2}
\begin{figure}[ht!]
\centering
\subfigure[$\,\,\,|\phi^+\rangle=\frac{1}{\sqrt{2}}(|0\rangle|0\rangle + |1\rangle |1\rangle)=\frac{1}{\sqrt{2}}(|+\rangle|+\rangle + |-\rangle |-\rangle)$\label{entangled}]{\includegraphics[width=3.4in]{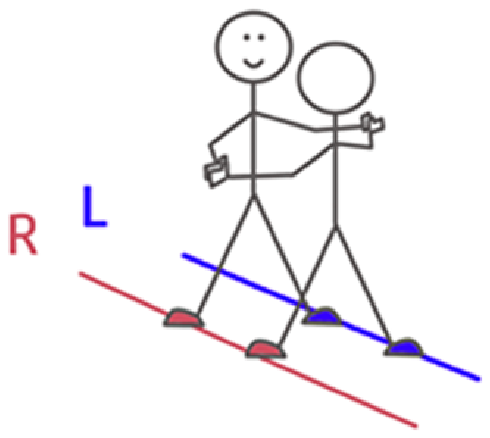}}
\subfigure[$\,\,\,|\phi^+\rangle \rightarrow |0\rangle|0\rangle$\label{entangledmeas}]{\includegraphics[width=3.4in]{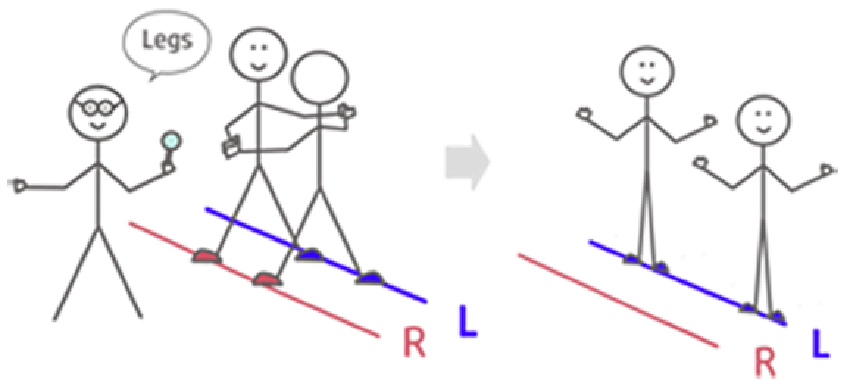}}
\caption{(a) State of two particles that are entangled ($|\phi^+\rangle=(|0\rangle|0\rangle + |1\rangle |1\rangle)/\sqrt{2}=(|+\rangle|+\rangle + |-\rangle |-\rangle)/\sqrt{2}$). (b) Measurement of the \textit{leg} property ($z$-measurement) of an entangled state. Both particles have to do the same --in this example, jump either to the left or to the right line--. }
\end{figure}

In this part, the scientists prepare pairs of entangled particles. In the entangled state, the particles are facing each other, holding arms and standing with their legs on different lines (see Fig.~\ref{entangled}). Note that now, the particles react randomly to a measurement of both leg and arm properties, since they stand with both separated legs and arms. That is, if the first particle is measured, the particle changes the state and announces a result accordingly (see rules for part 1). But now there is a new rule for the particles: the second particle is entangled to the first, so it also changes its state in the same way as its other mate-particle did, without being measured (see Fig.~\ref{entangledmeas}). Due to this first measurement, the entanglement between the two particles is broken --they no longer hold hands--. If the second particle is subsequently measured, it behaves according to the rules specified in part 1.

The game proceeds in the same way as before: the two particles emerge in an entangled state and move along the lines. Once one particle is hit by a scientist, the scientist chooses to measure either the leg or the arm property. Then also the other entangled particle can be measured by another scientist. Again, the role of the scientists is to figure out rules of the behavior, make predictions and, in particular, announce if they find out something special about the behavior of the two particles. This behavior is even more counter intuitive if one considers entanglement between particles that are far away. This can be represented by two long strings or ropes both particles hold. If one of the particles is measured, the other one still behaves in the same way as the measured one. The situation is the same as the one explained in this section, but now the entangled particles are far apart. Students should encounter that there are strong, non-local correlations between measurement outcomes that are not possible in a classical system.

\subsection{Part 3: Decoherence}\label{part 3}
We now move on to a more realistic situation where particles are not perfectly isolated, but interact with other particles from the environment. Usually scientists work hard in their laboratory to prevent this, because quantum features disappear due to such decoherence effects.

In this part, a third group of students is needed to play the role of environment particles.

\begin{figure*}[ht!]
\centering
\subfigure[\label{1partenv}]{\includegraphics[width=3in]{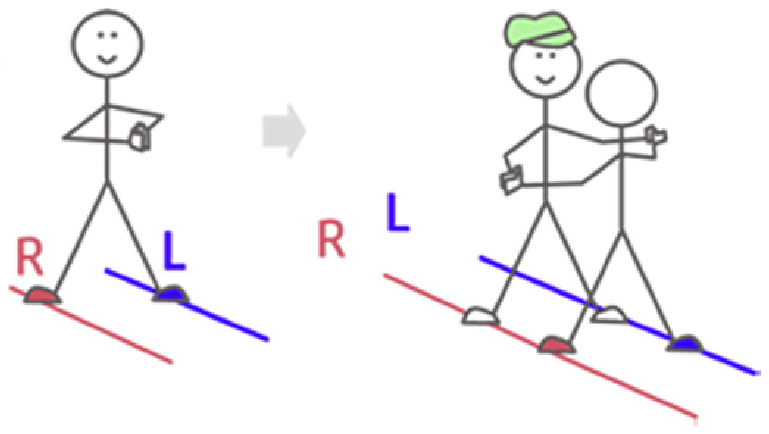}}
\subfigure[\label{2partenv}]{\includegraphics[width=3in]{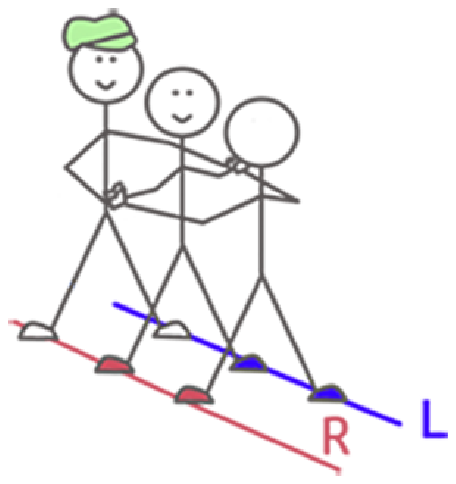}}
\subfigure[$\,\,\,\frac{1}{\sqrt{2}}(|0\rangle|0\rangle|0\rangle+|1\rangle|1\rangle|1\rangle)\rightarrow |0\rangle|0\rangle|0\rangle$\label{envmeasleg}]{\includegraphics[width=3in]{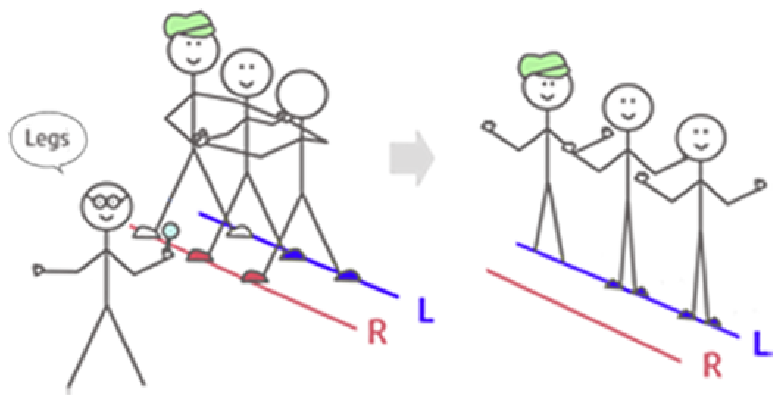}}
\subfigure[$\,\,\,\frac{1}{\sqrt{2}}(|+\rangle_1 \otimes |\phi^+\rangle_{2,E}+|-\rangle_1 \otimes |\phi^-\rangle_{2,E}) \rightarrow |+\rangle_1 \otimes |\phi^+\rangle_{2,E}$. \label{envmeasarm}]{\includegraphics[width=3in]{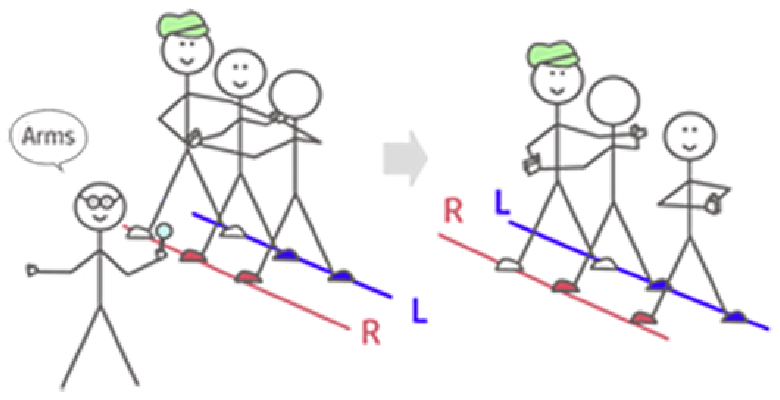}}
\caption{(a) A single particle in a state $|+\rangle$ gets entangled with a particle from the environment (represented by a stick figure with a cap in the picture). (b) One of the particles of the entangled pair interacts with the environment. (c) Measurement of the \textit{leg} property of a particle from the entangled state that has interacted with the environment. (d) Measurement of the \textit{arm} property of a particle from the  entangled state that has interacted with the environment. From figures (c) and (d), one can see that the measurement outcomes of the initially entangled pair are no longer correlated in both the \textit{arm} and the \textit{leg} properties, i.e. the interaction with the environment has broken the entanglement. See also the model of the interaction.\cite{model}}
\end{figure*}

We first consider a single particle that is prepared in the state $|+\rangle$ (\textit{arms front}). When it moves around, it eventually collides with another particle from the environment (represented with a cap in the pictures) and the two get entangled, i.e. they are in an entangled state as considered in part 2 (see Fig.~\ref{1partenv}). This process can be played as a catch-the-particle game, where the particles have to avoid being caught by the environment. The environment-particles are not under control of the scientists and cannot be measured. The particle follows the same rules as before when measured. Note however, that once the environment gets entangled to it, the particle has both legs and arms separated, i.e. scientists obtain random outcomes for all measurements. Thus, the particle that was in a pure state ($|+\rangle$), is in a mixed state when it interacts with the environment (see Sec.~\ref{theorback}).

In a similar way, one can investigate what happens to initially entangled states when one of the two particles collides with an environment-particle. Then, the entanglement between the initial particles is broken and all three particles together are now in an entangled state (see Fig.~\ref{2partenv} and a more detailed explanation of the model\cite{model} we use for describing the interaction with the environment).

Due to the interaction with the environment, the scientists obtain very different results from the previous entangled case. If the leg property is measured on one particle, the particle randomly chooses to jump either to the left or the right line, and the other particle has to jump to the same state, as it is shown in Fig.~\ref{envmeasleg}. From this point forward, the environment particle is also disentangled and can move away. If the arm property is measured, the particle randomly chooses to change either to the front or back state, while the other particle and the environment particle remain entangled (Fig.~\ref{envmeasarm}). Hence, if the second particle is subsequently measured, one finds a random outcome (no longer correlated to the result of the first particle). Therefore, the initially entangled pair of particles, that showed correlated results when either the leg or the arm property was measured, now shows non-correlated results when the arm property is measured. The entanglement has been broken due to the interaction with the environment.

\section{Quantum Cryptography}\label{QCryp}
The game presented in the previous section can be further used to work with more applications of quantum technology, e.g. quantum cryptography. \cite{Gis07,Gis02} Protocols like the BB84 for quantum cryptography rely on the principles of Quantum Mechanics described in Part 1 of the game, that is, quantum superposition states and measurements in different bases. The goal of such protocols is to establish a secret key, i.e. a random sequence of bits, only known to the sender, Alice, and the receiver, Bob. With help of this random key, an arbitrary message can then be reliably encrypted.

In the BB84 protocol, Alice randomly sends one of the four states $\{|0\rangle,|1\rangle\}$ (eigenstates of $z$-basis),$\{|+\rangle,|-\rangle\}$ (eigenstates of $x$-basis) to Bob, who measures the state in a randomly chosen basis (either $z$ or $x$). This is repeated $N$ times. The preparation (Alice) and measurement (Bob) bases are then announced publicly (not the measured values!), and only the set of bit values obtained when the two bases coincided are used for creating the key. 

Since only two bases ($x$ and $z$-bases) are needed, this protocol can be played with only part 1 of the game. The states Alice sends are simply $\{$legs left, legs right, arms front, arms back$\}$, and the corresponding bases are the \textit{leg} and the \textit{arm} properties. With this correspondence and the measurement rules specified in Sec.~\ref{part 1}, the BB84 protocol can be played directly. One student can play the role of Alice and another one the role of Bob. The rest of the students play the role of quantum particles. Further extensions can be made to include the role of an eavesdropper, Eve, that wants to intercept the secret key and get access to the message.

\section{Entangle me! In class}\label{class}
In this section, we present the results obtained from the experience we had testing the game in class. First, we give a brief description of how the game was played, and then we present the results of a brief survey given to the students after the session.

\subsection{The game}
The game was tested in May 2018 with the three science classes of 1 Bachillerato (a total of 73 high school juniors --16-year students--) at Colegio JOYFE in Madrid. The students had no previous background on advanced algebra and quantum mechanics. The session was divided into two slots of one hour each: the first one for playing the game and the second one for the discussion of the results and a brief presentation in which we explained the concepts of Quantum Mechanics and their relation to the game. The game was played outside, in the courtyard of the school, within a space of approx. $20-30 m^2$ (19-30 students). The lines on the floor were made of colored adhesive tapes. The class was divided in two groups --particles and scientists--, with more scientists than particles (approx. 2:1). The instructions were given to each group separately, so that the scientists do not know the particles' rules. The scientists were also given a sheet of paper for writing down the results, with a table of the form: ``Preparation state", ``Limbs you measured", ``Measured state". Due to the time resources we had, only part 1 --Single Particles-- (Sec.~\ref{part 1}) and part 2 --Entanglement-- (Sec.~\ref{part 2}) of the game were played. The specific instructions for the entangled particles in part 2 were given just after part 1 was finished, to make it easier for the particle-students to remember the rules.

The game was played as described in Sec.~\ref{implementation}. The scientists were arranged in two rows, from where they threw the balls at the particles. Every time they hit a particle (measurement), they wrote down the result and went to the end of the row to wait for the next turn.

Regarding the particles, in part 1, a group of them entered the lines in the state prepared by the scientists. Every time one particle was measured, it had to leave the lines. Once all the particles of the first round were out, the next round of  particles entered. In part 2, only one pair of entangled particles entered the lines. When this pair was measured, the following pair entered.

In part 2, we proposed the following exercise to the scientists: ``Collaborate in pairs, each of you measures one of the two particles of the entangled pair and writes down the results. Make as many combinations of measurements as possible so that you can compare results with your colleague afterwards".

After the game, we discussed the results in groups, so that the students could organize and interpret all the results together and share opinions. First, the students were asked if they had found any unusual behavior in the measurement results. If needed, more specific questions can be made to guide the discussion, such as: ``How were the results when you prepared \textit{legs} and measure the leg property? And if you measured the arm property?", ``In part 2, compare your results with the ones obtained by your colleague; is the behavior you observe the same if you measured arms and legs or not?". Finally, the students were asked to propose possible explanations for the physics behind the results, as if they were in a scientific congress.

From this experience, we highly recommend to plan the session for small groups of students supervised by a teacher (around 15 students per supervisor), so that the game is more dynamic and the students can participate more in the discussion. In addition, we have observed that the discussion encourages the students to be critical when analyzing the results and to be creative when imagining possible explanations for the results. 

We concluded the discussion with a brief presentation that highlighted the key concepts of quantum mechanics that can be learnt with a spin first approach, \cite{Dur13,Dur14,Dur16} to help the students understand the relation of their findings to these central concepts. We emphasized that quantum superpositions, stochastic behavior and state change under measurement, as well as Heisenberg's uncertainty relation, were illustrated in part 1 of the game, while the central concept of entanglement was illustrated in part 2.  

\subsection{Feedback from students}
In this section, we summarize the feedback of the students we collected after the session. We carried out an opinion survey to test the students' perception of the game, in order to evaluate the main difficulties they had and their suggestions for improvements.

The students were presented with a set of sentences that they could evaluate from 1 to 4, where 1 corresponds to full disagreement and 4 to full agreement. 

\begin{table*}[tbp]
\begin{center}
\begin{tabular}{|l|C{1cm}|C{1cm}|C{1cm}|C{1cm}|}
\hline
  &\textbf{1}&\textbf{2}&\textbf{3}&\textbf{4}\\
\hline \hline
The instructions were complicated & 60$\%$&32$\%$&4$\%$&4$\%$ \\ \hline
The game was easy to perform & 1$\%$&6$\%$&41$\%$&52$\%$ \\ \hline
I have enjoyed the game & 0$\%$&6$\%$&30$\%$&64$\%$ \\ \hline
I find the game useful to understand the new concepts & 0$\%$&15$\%$&32$\%$&53$\%$ \\ \hline
\end{tabular}
\caption{Results of the opinion survey given to the students after the game. A total of $n=73$ students were surveyed. The evaluation of each sentence goes from 1 (I fully disagree) to 4 (I fully agree).}
\label{tabla:results}
\end{center}
\end{table*}

The survey results presented in Table~\ref{tabla:results} show a good acceptance by the students. In addition, when asked for difficulties they had during the game, some of the scientists reported that they were too far from the particles so they could not hit them enough times to perform all the measurements they wanted to. The students were also asked for suggestions to improve the game. The most frequent suggestions were to have more time to get more results, and to make it more dynamic so that the scientists do not have to wait long for their turn. We consider that these suggestions could be fulfilled by having less students per game (we worked with groups of 28, 26 and 19 students in a limited space), so that each student can play more often.

With these results, we can conclude that the game can be done successfully, and that students do not find it complicated. In addition, it follows from the received responses that the students have fun with the game, which was also a crucial goal for us.

\section{Summary}
To summarize, we have introduced a game with simple rules that allows one to illustrate the basic principles of quantum mechanics, and to directly experience them. In addition, students can act and work like a real scientist, develop theories and test them. This should not only allow students to better remember rules and features of quantum systems, but also to grasp their significance and differences to classical systems. We have also shown that advanced quantum features such as entanglement and decoherence can in principle be illustrated in the same way, and even modern applications such as quantum cryptography can be treated.

\begin{acknowledgments}

We gratefully acknowledge Stefan Heusler for his useful comments, Mario Rodr\'iguez for making it real and Colegio JOYFE for letting us test the game with all the 16-year science classes. Thanks to all the students who collaborated, you did impressively well. This work was supported by the Austrian Science Fund (FWF) through projects P28000-N27 and SFB F40-FoQus F4012-N16.

\end{acknowledgments}


\begin{thebibliography}{26}
\bibitem{eigenstates} States $|0\rangle$ and $|1\rangle$ are eigenstates of the $z$-basis, whereas states $|+\rangle$ and $|-\rangle$ are eigenstates of the $x$-basis.
\bibitem{tensor} Another example of entangled state is the $|\phi^-\rangle=(|0\rangle|0\rangle - |1\rangle |1\rangle)/\sqrt{2}$ , which will be later used in this work as well. We omit tensor products for simplicity.
\bibitem{rules} If the students already have some background knowledge about Quantum Mechanics or if the teachers find it more suitable for their purposes, it is also possible to just tell the rules of both scientists and particles to all the students at the beginning of the game, so that they can experience both roles.
\bibitem{alternative rep} The body positions of the states may resemble the Bloch vectors of the states. The z axis in the Bloch sphere can be thought of as the left-right axis and the x axis as the front-back axis. So far, the representation introduced by the body positions does not account for the phase difference between $|+\rangle+|-\rangle$ and $|+\rangle-|-\rangle$ explicitly in the arms representation, only by the leg position ($|0\rangle=\frac{1}{\sqrt{2}}(|+\rangle+|-\rangle)$ (legs left) and $|1\rangle=\frac{1}{\sqrt{2}}(|+\rangle-|-\rangle)$ (legs right)). If more technical considerations want to be taught, one can also distinguish these states by facing different directions (one direction will correspond to a plus sign, and the other to a minus sign). However, we consider that this increases the complexity of the game and leads to difficulties in part 2 and part 3 of the game (e.g. it leads to the entangled state $|\phi^-\rangle$ instead of $|\phi^+\rangle$).
\bibitem{model} The system particles are initially in a $|\phi^+\rangle$ state, and the environment particle is in $|0\rangle$ state. One can use a simple model for the system-environment interaction and assume that the second system particle and the environment particle interact via a CNOT operation of the form: $|0\rangle|0\rangle_E\rightarrow|0\rangle|0\rangle_E$; $|1\rangle|0\rangle_E\rightarrow|1\rangle|1\rangle_E$. This yields a GHZ state $(|000\rangle + |111\rangle)/\sqrt{2}$, which can equivalently be written as $\frac{1}{\sqrt{2}}(|+\rangle_1 \otimes |\phi^+\rangle_{2,E}+|-\rangle_1 \otimes |\phi^-\rangle_{2,E})=(|+\rangle_1 \otimes (|+\rangle_2|+\rangle_E+|-\rangle_2|-\rangle_E)+|-\rangle_1 \otimes (|+\rangle_2|-\rangle_E+|-\rangle_2|+\rangle_E))/\sqrt{2}$. Measurement results of the system particles in the $z$-basis (legs) are random and perfectly correlated, while $x$-measurements (arms) are random for both particles and thus uncorrelated.
\end{thebibliography}
\end{document}